# Improving Scientific Article Visibility by Neural Title Simplification

Alexander Shvets[1][0000-0002-8370-2109]

[1] Universitat Pompeu Fabra, Barcelona 08018, Spain
`alexander.shvets@upf.edu`

**Abstract.** The rapidly growing amount of data that scientific content providers should deliver to a user makes them create effective recommendation tools. A title of an article is often the only shown element to attract people's attention. We offer an approach to automatic generating titles with various levels of informativeness to benefit from different categories of users. Statistics from ResearchGate used to bias train datasets and specially designed post-processing step applied to neural sequence-to-sequence models allow reaching the desired variety of simplified titles to gain a trade-off between the attractiveness and transparency of recommendation.

## 1 Introduction

The amount of information scientific society produces on a daily basis results in the necessity of researchers to have proper guidance in a digital space. The function of the virtual assistance is performed by various scientometric systems, research paper recommender systems (Haruna et al., 2017) and different kinds of search engines. (Shvets et al., 2015) summarizes the most common types of systems for scientometric analysis. The recent trend in scientific paper delivery is purpose specific web-resources, blogs, and e-journals often coupled with email subscriptions. They often provide personalized recommendations based on users' behavior and preferences.

The recommendation usually has a form of imprint often limited only by a title (as in the case with email subscriptions associated with limited space and lack of time to attract people's attention). Eventually, the success of recommendation depends on the informativeness of the title of an article subject to user's intentions and acknowledgment with a certain scientific field. This denotes the necessity of finding a way of varying the title of the same paper for different categories of users.

The focus of this paper is in developing models for creating a variety of simplified versions of the titles of scientific articles which would be condensed and informative enough and at the same time would correspond to the original topic of a paper to maintain users' loyalty. We aim at supporting two scenarios of personalized simplification: the first ensuring narrow focus on specific scientific concepts for goal-oriented experts and the second providing a general overview for researchers working on the edge of a topic willing to expand their horizons. The latter case should not be treated as a generation of clickbaits (catchy short misleading headings) that are to be blocked with the use of efficient machine learning approaches (Biyani et al., 2016).

There is a variety of algorithms that could be used for title simplification which is a rapidly growing research area (Bouayad-Agha et al., 2009; Saggion et al., 2015; Guo et al., 2018). As long as the defined task is similar to text compression and abstractive summarization we made a choice towards encoder-decoder neural architectures (Nallapati et al., 2016, Nikolov et al., 2018).

The remainder of the paper is structured as follows. In Section 2, we propose a method for scientific title diversification and simplification. Section 3 is devoted to describing the datasets used for training. Section 4 denotes the experiment setup. Section 5 provides the results of numerical experiments. Section 6 is dedicated to human evaluation. In Section 7, finally, we discuss results and outline future work.

## 2   Method

Recent advances in natural machine translation (NMT) incite to solve the task in a supervised manner controlling the style of a title by conditioning training data. The method we propose comprises the following steps: a) selecting a subset from an *abstract-to-title* dataset to impose conditions that would force a model to generate hypotheses with desirable properties; b) training a sequence-to-sequence (seq2seq) model; c) applying a model to *title-to-title* generation; d) performing post-processing step to remove unnecessarily repeated tokens; e) filtering titles with improper structure. The remainder of this section describes each step in details.

To create titles of different styles for various categories of researchers several datasets should be used. The set of highly popular scientific titles may help to generate attractive headings for users with interests peripheral to the subject of a paper. The condition to have a multi-word noun phrase $NP_{mw}$ in a target text is to avoid producing overly shortened pointless titles. In case each training example contains a reference text $R_t$ and a target text $T_t$ that have *similar $NP_{mw}$* (at least two common terms), a model might learn to preserve the most important concepts from original titles needed by experts. Figure 1 shows the training example with similar $NP_{mw}$-s in $R_t$ and $T_t$.

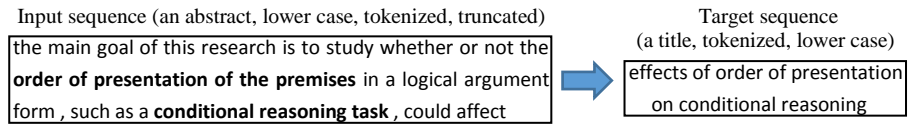

**Fig. 1.** Training example with similar noun phrases in reference and target text

We chose a bidirectional LSTM (Luong et al., 2015) with a copy mechanism (Gu et al., 2016; See et al., 2017) as a basic model. In particular, we used the realization in OpenNMT toolkit (Klein et al., 2018) enabling pointer that allows copying tokens from the reference text. The trained model is to be applied to new unseen titles, which are, in opposite to abstracts (cut-off after 50 tokens in our experiments), not truncated.

Since the task differs from general NMT task and summarization task by the absence of need in tracking alignment, traditional coverage mechanism (Wu et al., 2016), that discourage repetitions, is not included not to impose potentially harmful restrictions and not to overcomplicate the model. Instead, we introduce the post-processing step *PS* as follows. Firstly, each repetition of a term is removed leaving the only occurrence closest to the beginning of a text. Secondly, all the auxiliary tokens without required terms in between or after them are eliminated. In the end, we iteratively remove the last token in a text if it is an adjective or auxiliary token and, in addition, capitalize the title.

The last step consists in filtering improper titles, i.e., generated sequences that have less than two $NP_{mw}$-s similar to some $NP_{mw}$-s of the source title. In those use cases when even potentially pointless output is required, this step should be skipped.

## 3   Datasets

We chose ResearchGate[1] platform as a source of data. It has a recommender system and therefore openly counts the number of times a page with a paper was visited to provide reasonable recommendations that motivates authors to be more visible.

We selected 150K imprints of articles on various topics using a wide list of general scientific words (Osipov et al., 2014) as an entry point to the articles. Figure 2 shows the correlation between the number of paper views $N_v$ and the title lengths $L_t$ (in characters) in the collection. The top-viewed articles along negative correlation formed the desired set of highly popular titles. The whole pool of imprints formed a generic dataset. Random split for training and validation (93/7) was carried out. The set of 1000 imprints with $N_v = 1$ and $L_t > 100$ was used for testing the models.

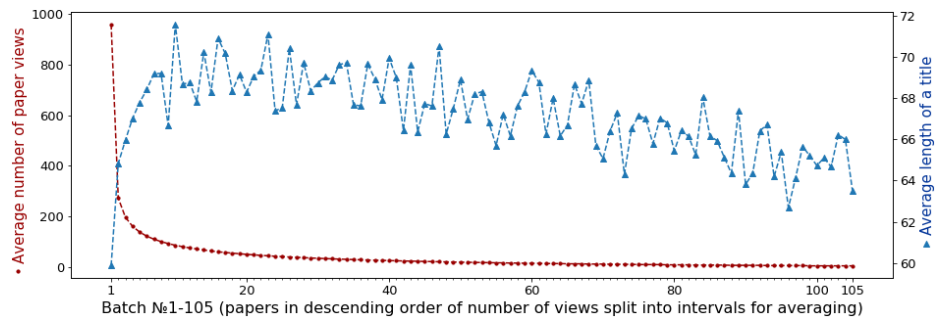

**Fig. 2.** Dynamics for titles in a professional social network ResearchGate ($N_v > 5$, $L_t > 20$).

The texts were pre-processed on the fly applying language detection with langid.py[2] and sentence detection with tokenization from NLTK[3]. Cleaning of training and vali-dation data consisted in leaving only examples with $N_v > 1$, at least one common term in $R_t$ and $T_t$, and $L_t > 20$.

---
[1]   https://www.researchgate.net/

To detect noun phrases we used Spacy chunker (Honnibal and Montani, 2017) that we elaborated for identifying complex phrases, which map single concepts (e.g., "vertex energy of a graph" that is a lexical variation of the concept "graph energy").

## 4 Experiment Setup

Selecting the first 75 characters of the reference text is generally used as a baseline in summarization tasks; cf., e.g., (Rush et al., 2015). We added subsequent cut-off after the last noun in a phrase. This improved baseline is referred to as $M_{Base}$ henceforth.

Several seq2seq models ($M_1$, $M_2$, ...) with the above-described architectures differed by a number of layers were applied to various datasets to bias the style of output text. They were then extended with post-processing PS ($M_{1ps}$, $M_{2ps}$, ...) and filtering steps, which are novel for the best knowledge of the author; cf. Table 1 for details.

**Table 1.** Distinctive details of basic and extended models

| Model | #layers | Dataset |
|---|---|---|
| $M_1 / M_{1ps}$ | 1 | conditioned ($R_t$ and $T_t$ have at least 2 pairs of *similar* $NP_{mw}$), 11K |
| $M_2 / M_{2ps}$ | 1 | strongly conditioned ($R_t$ and $T_t$ have at least 1 pair of *equal* $NP_{mw}$), 5.5K |
| $M_3 / M_{3ps}$ | 1 | weakly conditioned ($R_t$ and $T_t$ have a common term), 66K |
| $M_4 / M_{4ps}$ | 1 | *top-views* weakly conditioned ($R_t$ and $T_t$ have a common term), 18K |
| $M_5 / M_{5ps}$ | 2 | weakly conditioned ($R_t$ and $T_t$ have a common term), 66K |
| $M_6 / M_{6ps}$ | 2 | conditioned ($R_t$ and $T_t$ have at least 2 pairs of *similar* $NP_{mw}$), 11K |

For the final model assessment, we used measures BLEU (Papineni et al., 2002), ROUGE-1, ROUGE-2, ROUGE-L (Lin, 2004), and specially designed $NP_{diff}$-$p$, i.e., $NP_{mw}$-based precision evaluated as *rouge-L-p* considering a sequence of common $NP_{mw}$-s (only the first occurrence in a hypothesis for similar) regardless of their order.

The intermediate models created at checkpoints during the training were assessed, and the best by $NP_{diff}$-$p$ were selected as resulting.

## 5 Results

The most of the basic models performed reasonably: produced titles were in general shorter than original, multiple-word noun phrases from reference title covered a significant part of the generated title ($NP_{diff}$-$p$ = 0.68 on average). However, some models, especially $M_5$, introduced many repetitions (for all checkpoints): the BLEU value reflected it being equal to 0.18 for $M_5$ while the average value for the rest of models was equal to 0.35. Since BLEU depends on a number of same word occurrences, the increase of it by 24% on average due to *PS* attests usefulness of the step (cf. Table 2).

---

[2] https://github.com/saffsd/langid.py
[3] https://www.nltk.org

**Table 2.** Improvement of a title by post-processing step *PS*

| Original title (reference) | A Study on Knowledge Management System for Knowledge Competitiveness with One Stop Knowledge Service |
|---|---|
| Initial hypothesis before *PS* | knowledge management system for knowledge competitiveness with one stop knowledge service with one stop service with one stop service with one stop service with one stop service with one stop service with… |
| Resulting title | Knowledge Management System for Competitiveness with One Stop Service |

Filtering step allowed dropping less informative titles so that one can take advantage even of poor models reducing a risk to present misleading picking-eye headings or generic topics to an end user (cf. Figure 3 for examples of generated texts).

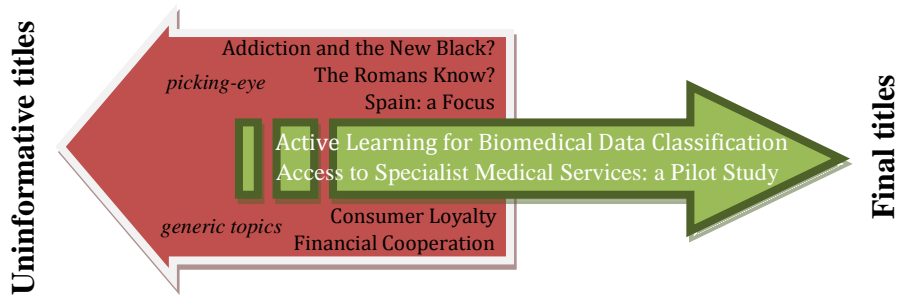

**Fig. 3.** Filtering step

The extension of basic models led to an increase of $NP_{diff}\text{-}p$ by 9% and *rouge-L-f* by 11% on average. Table 3 gives an idea of the variation of titles of different models in style and compression rate.

**Table 3.** ROUGE measures for inspected models

| model | rouge-1-r | rouge-1-p | rouge-1-f | rouge-2-r | rouge-2-p | rouge-2-f | rouge-L-r | rouge-L-p | rouge-L-f | rouge-L-f (basic $M_n$) |
|---|---|---|---|---|---|---|---|---|---|---|
| $M_{Base}$ | 0.60 | 1.00 | 0.74 | 0.54 | 1.00 | 0.69 | 0.60 | 1.00 | 0.66 | 0.64 |
| $M_{1ps+F}$ | 0.59 | 0.99 | 0.73 | 0.41 | 0.76 | 0.52 | 0.52 | 0.88 | 0.57 | 0.55 |
| $M_{2ps+F}$ | 0.58 | 0.98 | 0.72 | 0.42 | 0.78 | 0.53 | 0.53 | 0.89 | 0.58 | 0.53 |
| $M_{3ps+F}$ | 0.50 | 0.99 | 0.65 | 0.36 | 0.83 | 0.49 | 0.48 | 0.95 | 0.53 | 0.43 |
| $M_{4ps+F}$ | 0.52 | 1.00 | 0.67 | 0.34 | 0.75 | 0.46 | 0.47 | 0.89 | 0.52 | 0.43 |
| $M_{5ps+F}$ | **0.65** | 0.99 | **0.77** | **0.51** | 0.84 | **0.62** | 0.62 | 0.94 | **0.67** | **0.64** |
| $M_{6ps+F}$ | 0.50 | **1.00** | 0.65 | 0.38 | **0.89** | 0.52 | 0.48 | **0.96** | 0.52 | 0.48 |

It is worth noting that 1-layer models $M_1$ and $M_2$ trained on conditioned datasets reached higher values for the majority of measures in comparison to models $M_3$ and $M_4$ fed with generic data. This highlights rationality in pre-directing the training.

## 6 Human Evaluation

For human evaluation, we selected five papers of the NLP research group (TALN UPF) with titles longer than 93 characters (10-18 words). Their authors who own Ph.D. degrees were asked to rank output titles for these papers including original title by preference on clicking if they saw a title briefly in a daily email digest. To face different decision criteria assessors worked with papers of their authorship (for simulating expert behavior) and with papers of their colleagues (expanding horizons use case). If some titles in a set were the same or assessors did not have any preference between two similar titles they were allowed to rank them equally. The top models sorted by the average rank and examples of titles from one set are listed in Table 4.

**Table 4.** Top models according to the average rank given by assessors

| Model | Final Title | $R_{AVG}$ |
|---|---|---|
| $M_{Base}$ | Multisensor: Development of Multimedia Content Integration Technologies | 1.9 |
| $M_{3ps}$ | Multimedia Content Integration Technologies for Journalism | 3.7 |
| $M_{5ps}$ | Development of Multimedia Content Integration for Journalism, Media and International Exporting and Decision Support | 4.2 |
| $M_{Orig}$ | Multisensor: Development of multimedia content integration technologies for journalism, media monitoring and international exporting decision support | 4.3 |
| $M_{6ps}$ | Multimedia Content Integration Technologies for Journalism, Media | 4.4 |
| $M_{4ps}$ | Multimedia Content Integration for Journalism | 5.7 |

## 7 Discussion and Future Work

The noted final increase of $NP_{diff}$-$p$ and *rouge-L-f* indicates that common subsequences became longer in relation to the length of titles meaning that offered post-processing step with filtering plays an important role in forming a fluent text. At the same time, the output should not have been just one of the original subsequences; therefore, we did not aim at reaching too high precision values.

Pure state-of-the-art seq2seq models without post-processing step got low ranks on human evaluation. The models $M_{1ps}$ and $M_{2ps}$ have a higher average rank of 6. Their titles are well-formed and represent a combination of original multi-word expressions (cf. Table 3 for relatively high scores of *rouge-2-r*), however, less corresponding to the topic that is partly reflected by comparatively lower values of *rouge-L-p*. The outputs of the models $M_{3ps}$ and $M_{5ps}$ were often preferred to original titles. Having 1.3 times shorter titles than $M_{5ps}$, conditionally trained $M_{6ps}$ achieved almost the same average score. The baseline has the highest rank since it often better preserves the meaning although it does not always form a complete phrase. The main drawback is that it usually only generalize a title to some extent (in case of well-turned subsequence) and miss details experts might need.

The close average ranks of models and *rouge-L-f* on the same level for all models denote an opportunity to overcome the general problem of lacking the variability in neural seq2seq generation. Different title styles give a possibility to reach a preferable trade-off between the conciseness of the title and its transparency.

For future work, we plan to gain value from methods of paraphrasing (Cao et al., 2017), advanced simplification (Zhang and Lapata, 2017; Štajner and Saggion, 2018) and surface realization for deep input representations (Belz et al., 2018) to obtain diverse semantically close outputs differ from text reformulated with mostly the same words. Fake-paper detecting (Byrne and Labbé, 2017) and assessing the quality of scientific texts (Shvets, 2015) will help to avoid training the models on misleading titles. Finally, pre-existing taxonomies (e.g., JEL codes in Economics, the ACM taxonomy in Computer Science, the Web of Science categories attached to journals), and meta information of papers such as authors' keywords or KeywordsPlus items inferred from the references cited (Garfield and Sher, 1993) are to be used for preselecting the most relevant concepts to bias the training.

## Acknowledgments

The presented work was supported by the European Commission under the contract numbers H2020-700024-RIA, H2020-700475-IA, H2020-779962-RIA, H2020-786731-RIA, and H2020-825079-RIA and by the Russian Foundation for Basic Research under the contract number 18-37-00198. Many thanks to the four anonymous reviewers for their valuable comments, and to the five postdoctoral researchers for their high responsiveness in the evaluation and insightful feedback.